\documentclass[twoside,twocolumn,9pt]{article}
\usepackage{extsizes}
\usepackage[super,sort&compress,comma]{natbib} 
\usepackage[version=3]{mhchem}
\usepackage[left=1.5cm, right=1.5cm, top=1.785cm, bottom=2.0cm]{geometry}
\usepackage{balance}
\usepackage{mathptmx}
\usepackage{sectsty}
\usepackage{graphicx} 
\usepackage{lastpage}
\usepackage[format=plain,justification=justified,singlelinecheck=false,font={stretch=1.125,small,sf},labelfont=bf,labelsep=space]{caption}
\usepackage{float}
\usepackage{fancyhdr}
\usepackage{fnpos}
\usepackage[english]{babel}
\addto{\captionsenglish}{%
  
}
\usepackage{array}
\usepackage{droidsans}
\usepackage{charter}
\usepackage[T1]{fontenc}
\usepackage[usenames,dvipsnames]{xcolor}
\usepackage{setspace}
\usepackage[compact]{titlesec}
\usepackage{hyperref}
\usepackage[final]{pdfpages}

\usepackage{epstopdf}

\definecolor{cream}{RGB}{222,217,201}

\begin{document}

\pagestyle{fancy}
\thispagestyle{plain}
\fancypagestyle{plain}{
\renewcommand{\headrulewidth}{0pt}
}

\makeFNbottom
\makeatletter
\renewcommand\LARGE{\@setfontsize\LARGE{15pt}{17}}
\renewcommand\Large{\@setfontsize\Large{12pt}{14}}
\renewcommand\large{\@setfontsize\large{10pt}{12}}
\renewcommand\footnotesize{\@setfontsize\footnotesize{7pt}{10}}
\makeatother

\renewcommand{\thefootnote}{\fnsymbol{footnote}}
\renewcommand\footnoterule{\vspace*{1pt}%
\color{cream}\hrule width 3.5in height 0.4pt \color{black}\vspace*{5pt}} 
\setcounter{secnumdepth}{5}

\makeatletter 
\renewcommand\@biblabel[1]{#1}            
\renewcommand\@makefntext[1]%
{\noindent\makebox[0pt][r]{\@thefnmark\,}#1}
\makeatother 
\renewcommand{\figurename}{\small{Fig.}~}
\sectionfont{\sffamily\Large}
\subsectionfont{\normalsize}
\subsubsectionfont{\bf}
\setstretch{1.125} 
\setlength{\skip\footins}{0.8cm}
\setlength{\footnotesep}{0.25cm}
\setlength{\jot}{10pt}
\titlespacing*{\section}{0pt}{4pt}{4pt}
\titlespacing*{\subsection}{0pt}{15pt}{1pt}

\fancyfoot{}
\fancyfoot[LO,RE]{\vspace{-7.1pt}\includegraphics[height=9pt]{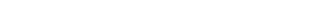}}
\fancyfoot[CO]{\vspace{-7.1pt}\hspace{13.2cm}\includegraphics{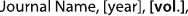}}
\fancyfoot[CE]{\vspace{-7.2pt}\hspace{-14.2cm}\includegraphics{head_foot/RF}}
\fancyfoot[RO]{\footnotesize{\sffamily{1--\pageref{LastPage} ~\textbar  \hspace{2pt}\thepage}}}
\fancyfoot[LE]{\footnotesize{\sffamily{\thepage~\textbar\hspace{3.45cm} 1--\pageref{LastPage}}}}
\fancyhead{}
\renewcommand{\headrulewidth}{0pt} 
\renewcommand{\footrulewidth}{0pt}
\setlength{\arrayrulewidth}{1pt}
\setlength{\columnsep}{6.5mm}
\setlength\bibsep{1pt}

\makeatletter 
\newlength{\figrulesep} 
\setlength{\figrulesep}{0.5\textfloatsep} 

\newcommand{\topfigrule}{\vspace*{-1pt}%
\noindent{\color{cream}\rule[-\figrulesep]{\columnwidth}{1.5pt}} }

\newcommand{\botfigrule}{\vspace*{-2pt}%
\noindent{\color{cream}\rule[\figrulesep]{\columnwidth}{1.5pt}} }

\newcommand{\dblfigrule}{\vspace*{-1pt}%
\noindent{\color{cream}\rule[-\figrulesep]{\textwidth}{1.5pt}} }

\makeatother

\twocolumn[
  \begin{@twocolumnfalse}
{\includegraphics[height=30pt]{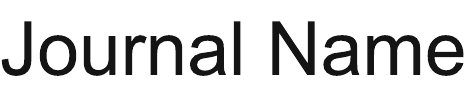}\hfill\raisebox{0pt}[0pt][0pt]{\includegraphics[height=55pt]{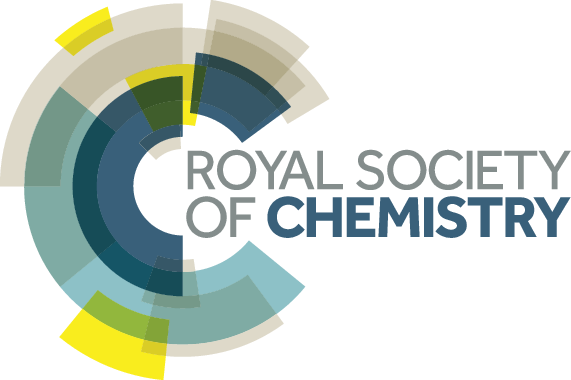}}\\[1ex]
\includegraphics[width=18.5cm]{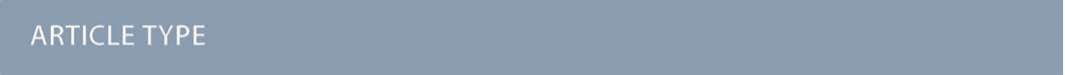}}\par
\vspace{1em}
\sffamily
\begin{tabular}{m{4.5cm} p{13.5cm} }

\includegraphics{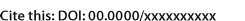} & \noindent\LARGE{\textbf{Pursuing colloidal diamond$^\dag$}} \\
\vspace{0.3cm} & \vspace{0.3cm} \\

 & \noindent\large{{\L}ukasz Baran,$^{\ast}$\textit{$^{a}$} Dariusz Tarasewicz,\textit{$^{a}$}, Daniel M. Kami{\'n}ski\textit{$^{b\ddag}$}, and Wojciech R{\.z}ysko\textit{$^{a}$}} \\

\includegraphics{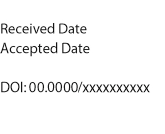} & \noindent\normalsize{The endeavor to selectively fabricate cubic diamond is challenging due to the
 formation of competing phases such as its hexagonal polymorph or others possessing similar free energy. 
 The necessity to achieve that is of paramount importance, since the cubic diamond 
 is the only polymorph exhibiting a complete photonic bandgap making it 
 a promising candidate in view of photonic applications.
 Herein, we demonstrate that due to the presence of external field
 and delicate manipulation of its strength we can attain the selectivity 
 in the formation of cubic diamond in a one-component system comprised of
 designer tetrahedral patchy particles. The driving force
 of such phenomena is the structure of the first adlayer which is 
 commensurate with (110) face of the cubic diamond. 
 Moreover, after a successful nucleation event, once the external 
 field is turned off, the structure remains stable, paving an avenue
 for a further post-synthetic treatment.} \\

\end{tabular}

 \end{@twocolumnfalse} \vspace{0.6cm}

  ]

\renewcommand*\rmdefault{bch}\normalfont\upshape
\rmfamily
\section*{}
\vspace{-1cm}


\footnotetext{\textit{$^{a}$~Department of Theoretical Chemistry, Institute of Chemical Sciences, Faculty of Chemistry, Maria-Curie-Sklodowska University in Lublin, Pl. M Curie-Sklodowskiej 3, 20-031 Lublin, Poland E-mail: lukasz.baran@mail.umcs.pl}}
\footnotetext{\textit{$^{b}$~Department of Organic and Crystalochemistry, Institute of Chemical Sciences, Faculty of Chemistry, Maria-Curie-Sklodowska University in Lublin, Pl. M Curie-Sklodowskiej 3, 20-031 Lublin, Poland }}

\footnotetext{\dag~Electronic Supplementary Information (ESI) available: See DOI: 00.0000/00000000.}


\section{Introduction}

Crystallization of hard spheres has been a matter of a long-standing
scientific discussion. It has been found     that the experimental
realization of such systems can be done by the application of
dispersed colloidal particles of, for instance,
polymethylmethacrylate dissolved in a non-polar solvent. \cite{Pusey1986, Cheng2001}
This result was particularly useful, as colloidal crystals have a size comparable to visible light, therefore making them particularly appealing in the context of photonic crystal synthesis. 
Photonic crystals or photonic band gap structures are three-dimensional periodic 
dielectric materials that are to photon waves as semiconductor crystals to electron waves
making them potentially useful for controlling the propagation of light. \cite{Yablonovitch:93} 
Such structures are omnipresent in nature, ranging from gem and opals formed
by precipitative settling processes to wing scales or barbules of some insects and bird feathers. \cite{phystoday1}
In view of the photonic applications, open lattices such as diamond \cite{photonic_appl} or tetrastack \cite{tetrastack1, tetrastack2}
serve to be great candidates since in both cases the
cubic polymorph exhibits a complete photonic band gap. 
Isotropic particles, however, possess the tendency to form the structures
of the highest packing fraction in which the stability is controlled
by the nearest neighbors with a similar local environment, resulting in 
the formation of two competing networks i.e. face-centered cubic (fcc)
and hexagonal close-packing (hcp) crystals. 

The endeavor to selectively fabricate colloidal cubic diamond is a formidable task. 
In recent years, a few routes have been utilized for the design of such tetrahedral open lattices. 
In this respect, patchy particles have been extensively used \cite{tetrastack2, sanz, sanz1, zwli, ZhanLi1,
Romano2012, Chakrabarti2018, Noya1, Noya2019, zhang1}.
The most straightforward approach is to distribute the patches over the spherical 
core of the particle matching the symmetry of the desired crystal structure. 
Although this seems to be a quite feasible way, tetrahedral patchy particles
exhibit a quite complex phase diagram which includes crystallization into
stacking hybrids of interlaced hexagonal (DH) and cubic diamonds (DC)  \cite{sanz, sanz1, Romano2012}.
Enforcement of the directionality of interparticle interactions
does not seem to alleviate these issues as the formation of empty clathrate cages
is now preferred \cite{Noya2019}.
All these obstacles result in that precise control over the nucleation process
becomes extremely challenging.
It is therefore of paramount importance to find directions for the selectivity 
in the synthesis of the target crystalline structure and to circumvent
the undesired coexistence with the second phase. 

Several routes have been utilized to alleviate the aforementioned issues. 
To list a few, the assembly of tetrahedral DNA origami constructs 
acting as host molecules caging guest gold nanoparticles coated with
single-stranded DNA was utilized to form a cubic diamond superlattice \cite{Ducrot2017}. 
Similarly, cooperative self-assembly of tetrahedra and isotropic spherical particles
lead to the formation of diamond and pyrochlore crystals \cite{Science2016}.
Recently, it has been demonstrated that the use of tetrahedral clusters with partially
embedded patches lead to the selective formation of the cubic diamond crystal
in a single-component systems \cite{He2020}. 
The idea lies behind the excluded volume effects which forbid the aggregation
of the building blocks into the eclipsed conformation, necessary for the
the emergence of the hexagonal diamond.

Patchy particles have also been extensively studied by the means of computer simulations. 
For instance, Romano and Sciortino \cite{Romano2012} demonstrated
that triblock patchy particles with spherical patches
crystallize simultaneously into two tetrastack polymorphs. 
However, patterning the patches into triangular shape 
results in the selective formation of a cubic tetrastack structure
or clathrates, depending on the relative orientation of the patches. 
{\color{black}The idea of patterning the patches is based on the seminal paper by Zhang et al. \cite{zhang1}
where they introduced additional potential energy term between pairs of bonded particles 
that induced relative orientations favoring staggered conformation. }
Similar studies have been performed for triblock patchy particles with
patches different in nature \cite{tetrastack2, Chakrabarti2018, Nano_rods2021}. 
Utilization of the so-called ``self-limiting'' approach in which
patchy particles form tetrahedral clusters which cease to grow further
under certain conditions and  instead act as a secondary building block.
Such a two-stage self-assembly route hinders the formation of different
undesired structures, partly due to the monodispersity attained in the first step of the assembly process. 
Sun \textit{et al.} \cite{zwli} also proposed an alternative route for the 
nucleation of colloidal photonic crystals which involved the utilization
of cooperative self-assembly of triblock patchy particles with 
spherical colloids, similar to the experimental realization \cite{Science2016}.
Binary mixtures of tetrahedral patchy particles have also been 
studied quite recently \cite{Tetraedr_PNAS, Rovigatti_mixture}.
It is also worth emphasizing that these studies, although performed \textit{in silico},
seem not to be 
challenging to be performed experimentally in view of the recent 
advances in the patchy particle synthesis \cite{He2020, Pawar2010, grzybowski_rev1, Duguet2017, Kim20, Rev2022}.

In spite of the recent successes, a comprehensive understanding 
of the conditions for the selective formation of colloidal diamonds
is yet to be achieved. 
Here we demonstrate another alternative route 
to attaining the cubic diamond polymorph
by precise control over the system parameters
by employing the bottom-up self-assembly strategy for tetrahedral
patchy particles in one-component system. 
Due to the presence of the external field and delicate manipulation of its strength the diamond networks start to nucleate
differently and the driving force of the process is the structure of
the first adlayer. The change in its structure leads to a change in the
nucleation mechanism and to the selective
emergence of the cubic diamond lattice. Moreover, once the external field
is turned off, the structure remains stable which gives an opportunity
for a further post-synthetic treatment.  

\section{Methods}

\textbf{Model details}

\begin{figure}[h!]
    \centering
    \includegraphics[width=0.5\textwidth]{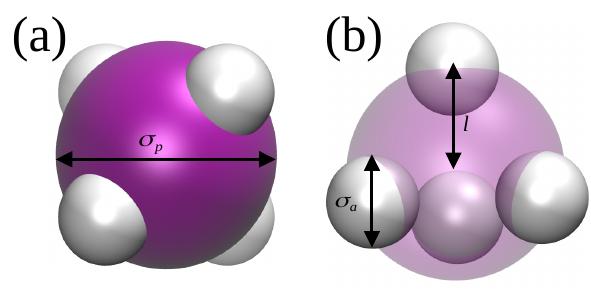}
    \caption{{\color{black}Schematic representation of the parameters of the model for tetrahedral patchy particles. 
     Central spherical core and active sites are schematically represented (scale is not preserved) by purple and white spheres, respectively. }}
    \label{fig:model}
\end{figure}

In this work, we employ a similar approach as already used in our previous papers
\cite{baran3, baran4}. 
The patchy particle model is comprised of the spherical core of the diameter $\sigma_p$ 
on which surface four attractive patches of size $\sigma_a$ are spatially distributed in tetrahedral arrangement and are embedded to a certain extent.
The parameter $l$ describing the latter depicts the joining distance between the centers of a core and each of the four patches {\color{black} (cf. Fig.~\ref{fig:model})}. 
We have shown that such an approach can be successfully used to mimic directional interactions. 
Moreover, by the precise control over the patch size and its embedment distance $l$,
one can attain a specific and desired patch's valency which we have proven to be nearly equivalent to the one used within the standard routine of Kern-Frenkel potential \cite{baran3, baran4, kaluzny2013}.
The patchy particles interacted via truncated and shifted Lennard-Jones (12,6) potential
in which there is no discontinuity in the forces. 
In the following, $\sigma_p=\sigma$ and $\varepsilon_{pp}$
have been defined to be the units of distance and energy, respectively. 
In the current simulations, we have chosen the set of parameters to be equal $\sigma_a=0.2\sigma$ and $l=0.36\sigma$ and have been devised to allow for the association of only pairs of patches resulting in that every patchy particle could create four bonds {\color{black} in total}. Moreover, we fixed the association energy to be $\varepsilon_{aa}=5.0\varepsilon$ whereas $\varepsilon_{pp}=1.0\varepsilon$. The range of interparticle interactions
was set to be $r_{cut,aa}=2.0\sigma_{aa}$ and $r_{cut,ij}=\sigma_{ij}$ otherwise ($ij=pp,pa$).
This indicates that the only attraction in the system was due to the association
between different patches and the remaining interactions were soft-repulsive. 
{\color{black}Embedding an active site to the certain extent which is controlled by the parameter $l$ results in that the spherical Lennard-Jones potential is screened for certain orientations of patchy particles. In effect, causing that one can control the directionality of the interparticle interactions.
Larger values of $l$ than $l=0.36\sigma$ lead to that one association site can 
create multiple bonds which was an undesired effect within the scope of 
current study (Supporting Information).}

{\color{black}In order to maintain the rigidity and the desired geometry of our tetrahedral particles, we used harmonic spring potentials for bonds and bond angles. The harmonic spring constants were set to be $k_b=1000 \varepsilon/\sigma^2$ and $k_\theta=1000\varepsilon/rad^2$  to ensure that the fluctuations in all the necessary bond and bond angles are negligible. }

On top of the interparticle interactions, we introduced the effects of the spatial constraints by the means of the flat, structureless wall modeled by the Lennard-Jones (9,3) potential is defined as follows:

\begin{equation}
 U^{\mathrm{ext}}(z)=\varepsilon_{wc} \left [\frac{2}{15} \left (\frac{\sigma}{z} \right)^9 - \left (\frac{\sigma}{z} \right)^3 \right ] 
 ~~~~~~~{\rm if~} z\leq5\sigma
\end{equation}

\noindent where $\varepsilon_{wc}$ indicates the depth of potential well for interactions.

The potential acted solely on the core of the patchy particle which on the one hand
should not interrupt the self-assembly behavior but on the other hand, 
it can be thought of as the introduction of isotropy on how the
patchy particles will be arranged with respect to the wall. 
Moreover, since the employed potential is uniform, it can be viewed as the presence of any external field and the change in the ratio $\xi$ of attraction strengths 
defined as $\xi=\frac{\varepsilon_{wc}}{\varepsilon_{aa}}$
should be universal within the range of the same exact phenomena. 

\noindent \textbf{Simulation protocol}

All of the molecular dynamics simulations have been launched by the use of 
LAMMPS simulation package \cite{LAMMPS}. Trajectories were evolved
using the velocity Verlet algorithm, with a timestep of $\tau=0.001$. 
The temperature has been maintained by the use of Nos{\'e}-Hoover chains
thermostat with damping factor $\tau_{NH}=10\tau$.
The systems comprised of $2500$ or $16129$ for systems under confinement 
to check for possible finite-size effects and the surface density was fixed to allow for the formation of seven close-packed monolayers. 
Therefore, the system size $L_x\times L_y\times L_z$ 
was equal to $18\times18\times40$ ($48\times48\times40$)
in $x, y, z$ directions for $2500$ (16129) patchy particles, respectively. 
Bulk systems were composed of $3375$ or 15625 patchy particles at distinct densities 
ranging from $\rho=0.2$ to $\rho=1.2$. 
Each of the systems was gradually cooled down from disordered states
with an increment in temperature equal to $\Delta T=0.01$.
Once the parameter $c_3(i,j)$ indicated the nucleation event, the
increment was changed to $\Delta T=0.005$. 
Simulations were launched for $2\times10^8-4\times10^8$ simulation steps at every thermodynamic state for the equilibration period. 
Further production runs were launched for at least $10^7$ timesteps
Five independent replicas have been launched in order to 
present reliable results and to calculate the error which are
presented in terms of standard errors.

\noindent \textbf{Block analysis}

The bulk phase diagram has been determined by using finite size scaling of the block density distribution functions following the method proposed by Binder \cite{Binder_bloki}.
The simulation cell at distinct densities ranging from $\rho=0.2$ to $\rho=1.2$,
has been divided into blocks of different sizes. Then, the density probability
distributions for patchy particles $P(\rho)$ normalized so that the integral equals unity,
have been estimated for each block.
An example of such a distribution function exhibiting double-peak behavior,
corresponding to the two-phase coexistence is shown in Supporting Information. 
The densities of coexisting phases have been taken from the maxima of these
distribution functions at every considered temperature.

\noindent\textbf{Crystal identification}

In order to quantify the formed three-dimensional
crystalline networks, we employed a standard routine involving
the calculation of Steinhardt order parameter \cite{steinhardt} given by:

\begin{equation}
 q_{l}=\sqrt{\frac{4\pi}{2l+1} \sum_{m=-l}^{m=l} \left| q_{lm} \right|^2}
 \label{eq:stein}
\end{equation}

\noindent with

\begin{equation}
 q_{lm}=\frac{1}{N_b(i)} \sum_{j \in N_b(i)} 
 Y_{lm}(\theta_{ij}, \phi_{ij}) 
 \label{eq:stein1}
\end{equation}

\noindent where $Y_{lm}$ are spherical harmonics and for a given sphere $i$, we choose a set of its 
nearest neighbors, $N_b(i)$. We define that any two spherical particles connected by a bond if they are neighbors, that is, if $ j \in N_b(i)$. For a particle $i$, the set of unit vectors $\mathbf{n}_{ij}$ points from $i$ to the particle $j$ $\in N_b(i)$ in the neighborhood of $i$. 
Each vector $\mathbf{n}_{ij}$ is characterized by its angles in spherical coordinates $\theta_{ij}$ and $\phi_{ij}$ on the unit sphere, evaluated
between the bond and an arbitrary but fixed reference frame.
The set of all bond vectors is called the bond network.

Discrimination between diamond networks is based on
the correlation function $c_l(i,j)$ \cite{chill+} defined as:

\begin{equation}
 c_l(i,j)=\frac{\displaystyle \sum_{m=-l}^{m=l} q_{lm}(i)q^*_{lm}(j)}{\left( \displaystyle\sum_{m=-l}^{m=l} q_{lm}(i)q^*_{lm}(i)\right)^{1/2}\left(\displaystyle \sum_{m=-l}^{m=l} q_{lm}(j)q^*_{lm}(j)\right)^{1/2}} 
 \label{eq:chill}
\end{equation}

Particles in a crystalline environment were labeled using $c_3(i,j)$. 
Two particles are connected by a staggered bond when $c_3(i, j) \leq -0.8$ and an eclipsed bond when $-0.05 \geq c_3(i, j) \geq -0.2$.
Afterward, molecules with exactly four neighbors were 
further discriminated as cubic diamond 
if they were connected with one another
by four staggered bonds or hexagonal diamond which
has three staggered bonds and one eclipsed bond.

\noindent\textbf{Characterization of the primary adlayer}

Another quantity that allows one to identify two-dimensional ordering,
including discrimination between the hexagonal and
Kagom\'e structures (and other) has been proposed quite recently \cite{Eslami2018, Eslami20181}.
It is defined as follows:

\begin{equation}
 \lambda_1(i)=\frac{1}{N_b(i)}\sum_{j \in N_b(i)}
 \left [ \sum_{m=-6}^6 \hat{q}_{6m} \hat{q}^*_{6m}
 - \sum_{m=-4}^4 \hat{q}_{4m} \hat{q}^*_{4m}\right ]
\end{equation}

\noindent and

\begin{equation}
 \lambda_2(i)=\frac{1}{N_b(i)}\sum_{j \in N_b(i)}
  \left [ \sum_{m=\pm6,\pm4} \hat{q}_{6m} \hat{q}^*_{6m}
 - \sum_{m=\pm4} \hat{q}_{4m} \hat{q}^*_{4m}\right ]
\end{equation}

\noindent where 

\begin{equation}
    \hat{q}_{lm}=\frac{q_{lm}(i)}
    {\left ( \displaystyle \sum_{m=-l}^l \left |q_{lm}(i) \right |^2 
    \right)^{1/2}}
\end{equation}

The hexagonal network is identified in the order parameter space $(\lambda_1,\lambda_2)$
with the values of $(0.0,0.8)$. Note that these values correspond to the perfect lattices
and to account for that, we allowed for $10\%$ uncertainty during the discrimination procedure. The degree of the order has been further evaluated as the ratio
of the number of patchy particles in the hexagonal environment $N_{\rm{hex}}$ and the total number of particles belonging to the first adlayer $N_{\rm{lay}}$, i.e.

\begin{equation}
    P_{\rm{hex}}=\frac{N_{\rm{hex}}}{N_{\rm{lay}}}
\end{equation}

\noindent\textbf{Calculated observables}

Aside from the aforementioned parameters, the calculation of density profiles $\rho(z)$, 
in the direction perpendicular to the wall has been performed.
This parameter supports the analysis and allows one to discriminate
the crystallographic direction of the growth of the formed solid network
and provides a hint at the spatial arrangement of the three-dimensional structure.

Another quantity, describing the surface behavior considered systems of interest, was the excess adsorption $\Gamma$ which is defined as follows

\begin{equation}
    \Gamma=\int_0^{\infty} (\rho(z)-\rho_b)dz
\end{equation}

\noindent with $\rho_b$ being the bulk density. In order to avoid
the effect of the presence of the second reflective wall, the upper
limit in the integral was taken as $L_z-4$.

To characterize the ``roughness'' of the adsorbed crystalline film,
we calculated the distribution of the film height. The simulation cell
in the xy-plane has been divided in the blocks of $1.0\sigma\times1.0\sigma$ size. 
In each of such bins, the maximum position value
in the z-direction of an atom belonging to the crystalline network
has been recursively evaluated. Then, the film height probability distribution
$P(h)$ is normalized so that the integral is equal to unity.

\section{Results}
\textbf{Bulk phase diagram} \\

\begin{figure*}[h!]
    \centering
    \includegraphics[width=\textwidth]{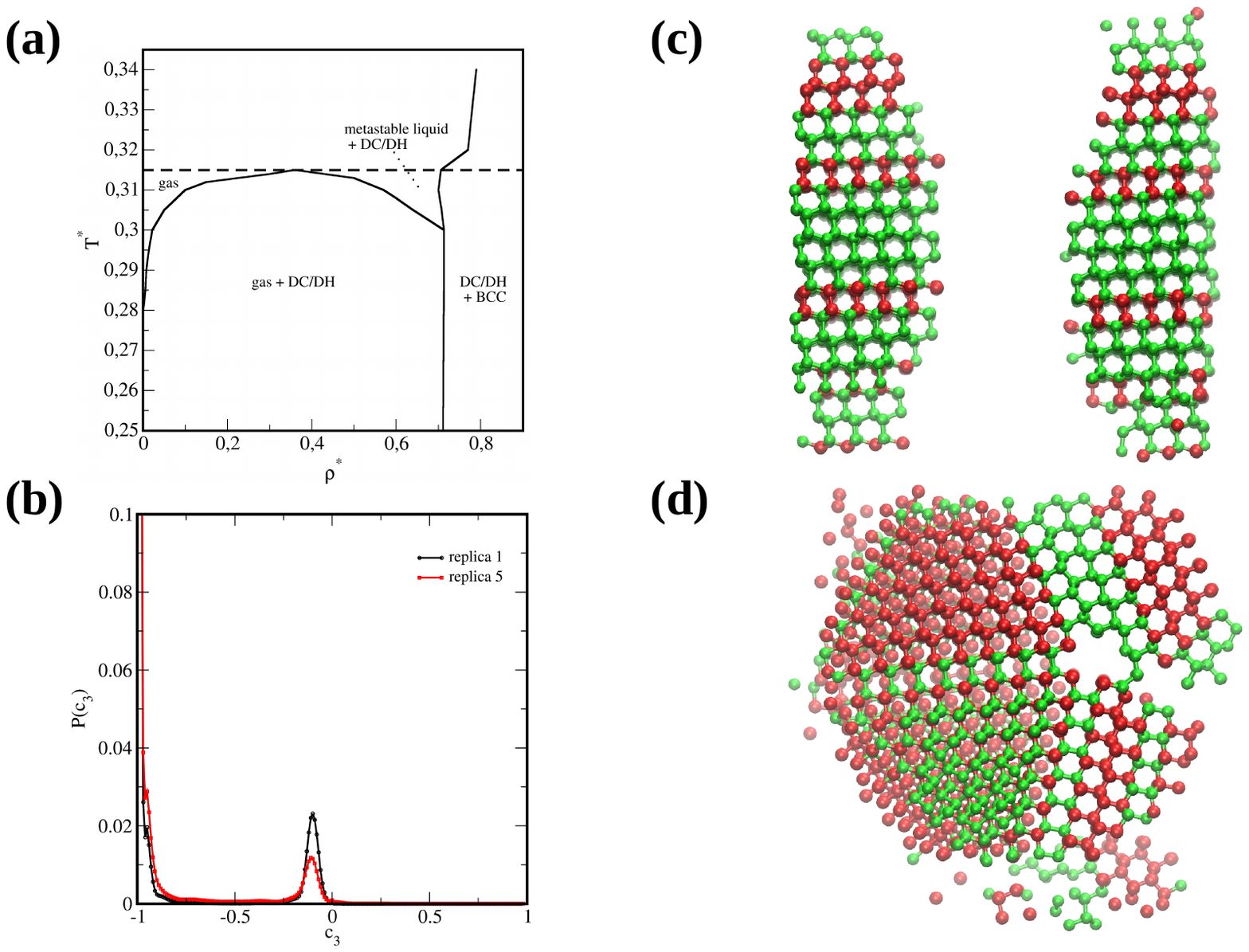}
    \caption{Part (a): Bulk phase diagram. 
    Part (b): Distribution functions
    of the order parameter $P(c_3)$ for two distinct replicas
    differing in the crystal composition. Parts (c), (d): Snapshots corresponding
    to the distribution functions shown in part (b). 
    Red and green sticks and atoms correspond to the cubic and hexagonal
    diamond environments, respectively.}
    \label{fig:bulk}
\end{figure*}

In order to accurately demonstrate the effect of the applied external field, we begin the discussion with the evaluation of the bulk phase diagram
for our designer tetrahedral patchy particles (cf. Fig.~\ref{fig:bulk} (a)). 
The patchy particles were devised such that 
the interactions resulted in a single association.
The evaluation of the bulk phase diagram was done by means of the block analysis proposed by Binder \cite{Binder_bloki}. An example of the distribution
function exhibiting double-peak behavior corresponding to 
the two-phase coexistence can be found in Supporting Information.

It is evident that the phase diagram, shown in Figure~\ref{fig:bulk} (a) 
is similar to the one reported by Romano \textit{et al.} \cite{sanz1} for the 
Kern-Frenkel model. In such a diagram, there is a gas-solid coexistence with a narrow 
(or not at all)  metastable liquid region. 
 A detailed description of how we envisaged the latter
can be found in Supporting Information.
This is particularly beneficial because the supercooled liquid phase usually has similar free energy to the crystal phases resulting in strong competition between the formation of kinetically arrested glasses and diamond crystals. We conjecture that the inhibition of the emergence of a liquid phase,
which is mainly due to the insufficient range of interparticle interactions, 
makes the crystallization process much more feasible. The latter conclusion
apart from the works by Romano \textit{et al.} \cite{sanz, sanz1}, 
can be supported by the study of Trokhymchuk \textit{et al.} \cite{Trokhymchuk}  which although
performed in 2D, suggests such behavior should be universal.

\begin{figure*}[t!]
    \centering
    \includegraphics[width=\textwidth]{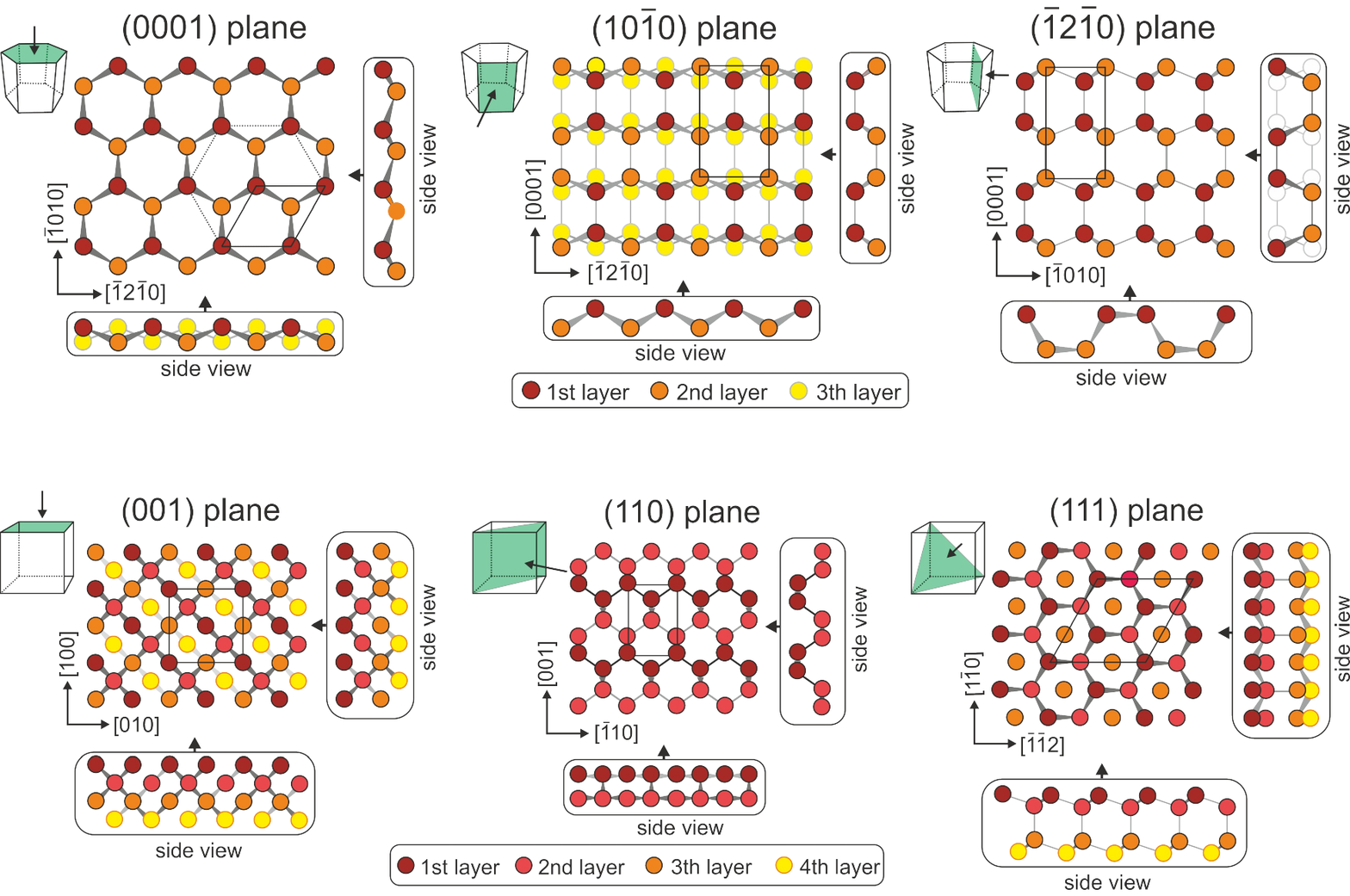}
    \caption{Schematic representation of symmetry faces of
    hexagonal (top panel) and cubic diamond lattices (bottom panel).
    The sketch is based on the one reported in ref~\cite{diamond_faces}.}
    \label{fig:faces}
\end{figure*}

However, similar to the previous studies, 
encountering the formation of stacking hybrids of cubic and hexagonal diamonds is inevitable. 
The discrimination between different local crystalline environments 
has been performed based on the $c_3(i,j)$ parameter.
Since these structures belong to different space groups being $Fd\bar{3}m$ and $P6_3/mmc$ for cubic and hexagonal polymorphs, respectively, such assembling can only be possible along the (111) direction of the former and (0001) of the latter (cf. Fig~\ref{fig:faces}).
Such a combination, although being described by a hexagonal unit cell, 
does not belong to any of the two space groups
and has a trigonal $P3m1$ space group, instead \cite{Stacking_water}. 

The value of order parameter is equal to $c_3(i,j)=-1$ and $c_3(i,j)=-0.1$ for staggered
and eclipsed bonds, respectively. 
Figure~\ref{fig:bulk} (b) demonstrates the distribution function $P(c_3)$ for
two replicas differing in the ratio of cubic and hexagonal diamonds.
As these structures differ by just one eclipsed bond,
the height of the distribution function around $c_3(i,j)=-0.1$
is different depending on the crystal composition. 
The representative snapshots corresponding to the distribution functions are shown
in parts (c, d) of Figure~\ref{fig:bulk}. 
It is worth noticing that the crystallization process in bulk 
systems changes from replica to replica and does not necessarily lead
to the formation of one single crystalline network (cf. Fig~\ref{fig:bulk} (c))
but can result in growth in several competing directions (cf. Fig~\ref{fig:bulk} (d)).
As will be shown later, the introduction of the directionality of the nucleation process may result in the prohibition of the formation of stacking hybrids. 
However, it is noteworthy that such interlaced stacking arrangements of cubic and hexagonal diamonds have also been found in other tetrahedral systems, such as water, diamond, or silver iodide to list a few \cite{Lupi2017, Stacking_water, Stacking_AgI, Murri2019}. \\

\noindent \textbf{External field-driven self-assembly} \\

\begin{figure*}
    \centering
    \includegraphics[width=\textwidth]{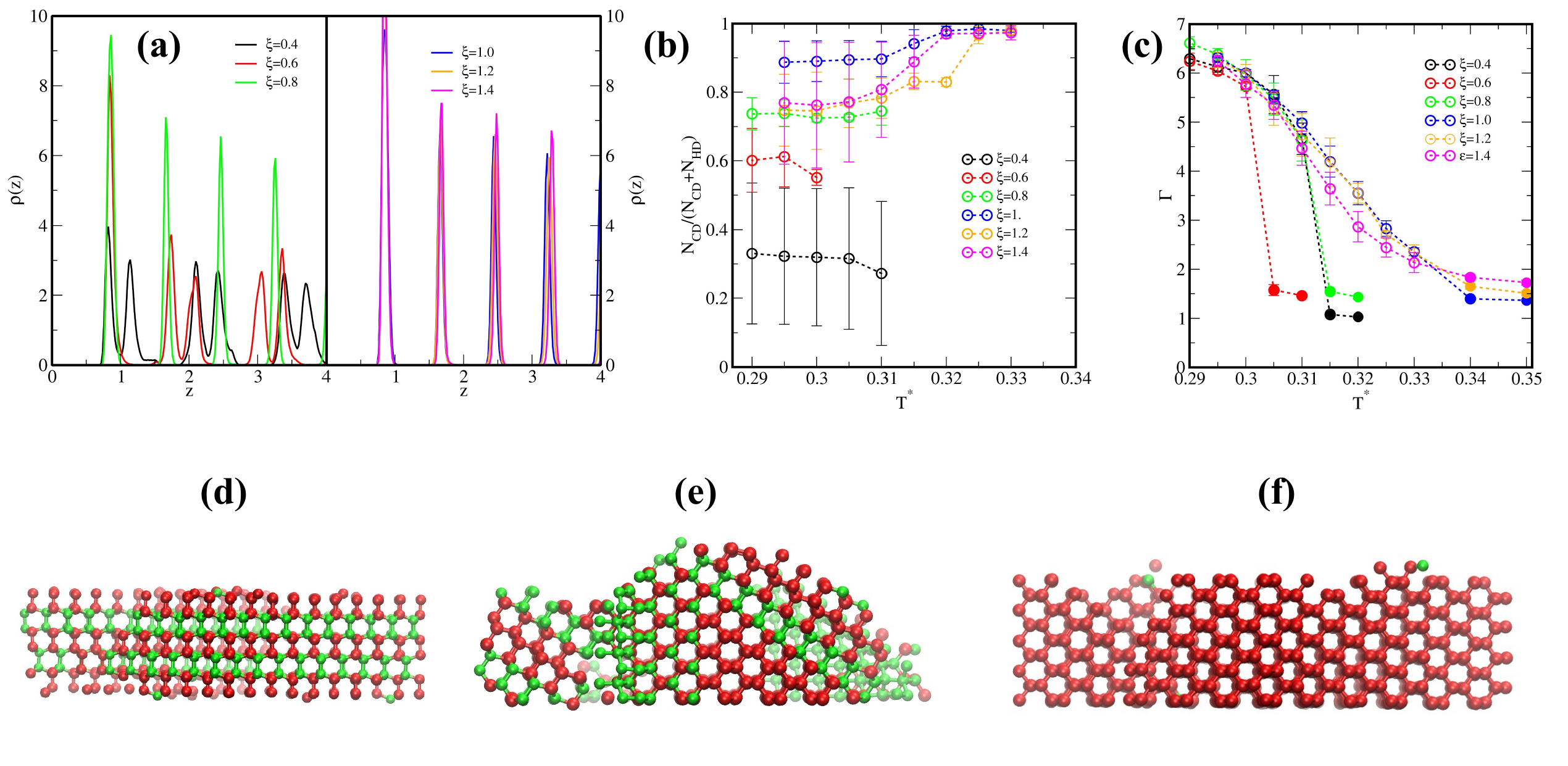}
    \caption{Part (a): Density profiles for considered systems
    under different strengths of an applied external field. Parts (b), (c): temperature
    relation of the cubicity parameter and excess adsorption for all considered systems. Parts (d)-(f): Snapshots demonstrating different crystal growth mechanisms
    depending on the strength of the surface potential; 
    $\xi=0.6$ (d), $\xi=0.8$ (e), $\xi=1.0$ (f). 
    Red and green sticks and atoms correspond to the cubic and hexagonal diamond environments, respectively.}
    \label{fig:external}
\end{figure*}

Relying on the evaluated bulk phase diagram, we can comprehensively assess 
the effect of the applied external field. The discussion will be based
on the introduction of the parameter $\xi$ depicting the ratio of 
the strengths of particle-wall and particle-particle energies.
We identified three different regimes on how the designer
patchy particles nucleate depending on the value of $\xi$. 
These findings are mainly supported by the density profiles $\rho(z)$ (cf. Fig.~\ref{fig:external} (a)), the order parameter $c_3(i,j)$ required for the discrimination between the diamond phases, and the excess adsorption curves (cf. Fig~\ref{fig:external} (c)). 

In the first case, $\xi=0.4$, indicating that the patch-patch interactions prevail, 
we observe the emergence of similar stacking hybrids as in the bulk phase (Supporting Information). 
A characteristic bilayer structure perpendicular to the z-axis can be seen,
indicating the (0001) or (111) faces of hexagonal and cubic diamonds, respectively.
The composition of the formed crystalline network has been described by the cubicity parameter defined as $N_{CD}/(N_{CD}+N_{HD})$ with $N_{CD}$ and $N_{HD}$ being the numbers of 
patchy particles belonging to cubic or hexagonal diamond networks, respectively, 
plotted in Fig.~\ref{fig:external} (b). 
Moreover, for such a small value of confining energy, we barely see 
any adsorption and the density profile resembles a bulk-like one. 
Large error bars in the cubicity curve over an entire temperature range also prove
the behavior to be similar to the bulk where the composition of the formed crystalline networks
is replica-dependent. 

Increasing the strength of the external field to
$\xi=0.6$ only slightly changes the behavior of the system. 
From the density profile, the emergence 
of the first layer differing significantly from the other part of the system
is evident (cf. Fig~\ref{fig:external} (a)).  
However, in spite of that, the first adlayer is not ordered, which has been
assessed by means of $(\lambda_1, \lambda_2)$ 2D order parameters introduced by M{\"u}ller-Plathe's group. 
It is also noteworthy that shifting the density profiles by $1/3\sigma$ in the z-axis,
leads to the overlap with the $\xi=0.4$ case, indicating similar behavior (see Supporting Information). This is also demonstrated by the cubicity parameter
whose value fluctuates around $0.6$ with quite large deviations. 
The overall structure resembles the one obtained in bulk or for $\xi=0.4$
which can be seen in Fig.~\ref{fig:external} (d).

A further increase of the confining energy to $\xi=0.8$ completely changes the behavior of the system. For the current case, we see that the characteristic bilayer structure
perpendicular to the z-axis is no longer present, indicating 
that the crystal growth mechanism changes. Moreover, the cubicity parameter
increases up to $0.8$. The snapshot displayed in Fig.~\ref{fig:external} (e) demonstrates 
that the formed crystalline network is completely different than in the previous cases.
The emergence of (110) face which is a characteristic plane for a cubic diamond 
can be observed. The imperfections 
are also present, however, their origin is due to the tilt in the crystal 
nucleus. In such cases, if the growth is tilted by $45^\circ$ the
(111) face becomes exposed and in consequence, the stacking hybrid
may form. 
The overall features presented for $\xi=0.8$ are even more pronounced for systems with $\xi \ge 1$. Complementary snapshots for different replicas in these systems are shown
in Supplementary Information. 

Since we are examining systems differing in the strength
of the confining energy, it is instructive to check how 
the adsorbed film grows on the surface. 
In order to do that, we calculated the excess adsorption
and its temperature variation is shown in Fig.~\ref{fig:external} (c). 
In general, it has been established that there are three
main scenarios on how the film can grow on solid substrates \cite{Pandit, Patrykiejew}. 
In the first one, when the surface adhesion dominates the cohesive interactions,  
the film grows asymptotically toward the infinite thickness
(Frank-van der Merwe: type-1). In the second mechanism, when the 
surface interactions are weaker, the film reaches a finite thickness usually only a few molecular layers thick (Stranski-Krastanov: type-2).
Eventually, when the interparticle interactions prevail, the adsorption is 
small under any conditions (Volmer-Weber: type-3). 

In currently investigated systems, we can see that the system with $\xi=0.4$ displays a type-3 behavior whereas $\xi=0.6$ and $\xi=0.8$ present type-2 behavior. 
All such observations can be confirmed by the inspection of the density 
profiles (Fig.~\ref{fig:external} (a)). 
A characteristic jump in the value of $\Gamma$ is observed for all such cases and further continuous growth is observed with a decrease in temperature. 
With further increase in the confining energy, we can see that for
$\xi \ge 1$ the growth mechanism changes to be type-1. 
In such instances, usually one should observe the epitaxial layer-by-layer growth
appearing as a series of layering transitions, 
however, we find a continuous increase of excess adsorption, instead.
We conjecture that the origin of such discrepancy is twofold. 
The simulation time and the cooling step can be insufficient for
these transitions to reveal. Moreover, as already mentioned,
for some replicas we observe that the crystal can be slightly tilted
which leads to the irregular structure of the surface. 
As we are characterizing the \textit{average} value of excess adsorption 
and not the \textit{distribution} in the xy-plane we are neglecting
the surface fluctuations which may also explain
the absence of layering transitions.
The aspect of the nature of surface phase transitions
should therefore be further studied. 

\begin{figure*}
    \centering
    \includegraphics[width=\textwidth]{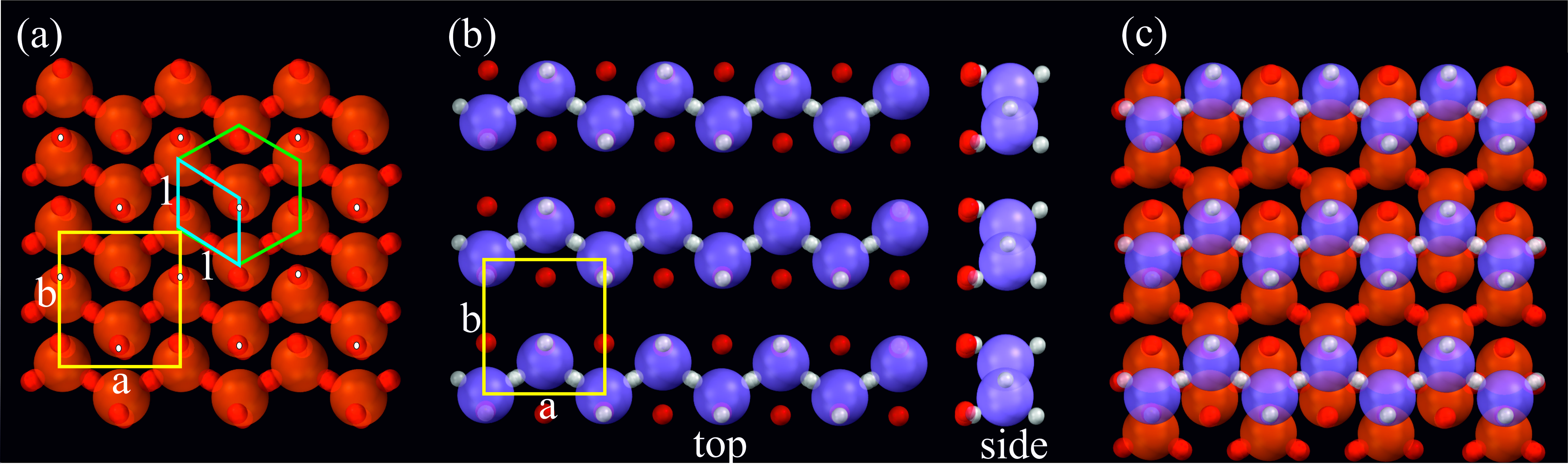}
    \caption{{\color{black}Part (a): Snapshot of the primary adlayer of hexagonal structure
    with the hexagonal (blue lines) and rectangular unit cells (yellow lines).
    $a=1.73\sigma$ and $b=2.0\sigma$. 
    Active sites forming bonds with the upper layer are marked with white dots. 
    Part (b):  Top and side view on the consecutive layer displaying the (110) face of cubic diamond which has the same vectors in its rectangular unit cell as the hexagonal adlayer from part (a). 
    Part (c): Snapshot showing how these two layers from parts (a, b)
    are spatially arranged one with another.
    White and red spheres schematically represent the active sites.
    Orange and purple atoms are used to pronounce the ordering and spatial
    arrangement.}}
    \label{fig:ucell}
\end{figure*}

Being acquainted with the behavior of the currently examined systems,
a naturally arising question can be cast: what is the driving force of such a reorientation resulting in the selective formation of the cubic diamond?
Why does the increase in the strength of the external field causes 
such a difference in the nucleation mechanism? Here, we aim
to answer that with the support of $(\lambda_1, \lambda_2)$ 
order parameter. This parameter is particularly
useful in terms of the characterization of two-dimensional 
networks as in the current case of the first adsorption layer.  
We already mentioned the fact that for $\xi=0.6$, despite
that there is a significant adsorption layer formed, it is not ordered. 
The first system in which the set of parameters 
$(\lambda_1, \lambda_2)$ identify the formation of the hexagonal network (up to 
$P_{\text{hex}}=0.2$) is $\xi=0.8$. Simultaneously, on the density profiles $\rho(z)$
we see that the crystal growth mechanism changes. Such behavior is even more explicit for $\xi \ge 1.0$ where $P_{\text{hex}}$ reaches values close to $0.9$
indicating significant ordering in the first adlayer. 

{\color{black} In order to demonstrate the mechanism governing the emergence of (110) face of cubic diamond, we refer to Figure~\ref{fig:ucell}. In panel (a), the primary adsorption layer of hexagonal structure is shown together with its hexagonal (blue lines) and rectangular unit cells (yellow lines). Part (b) of Figure~\ref{fig:ucell} displays the consecutive layer which is (110) face of cubic diamond which has identical rectangular unit cell as the hexagonal layer shown in panel (a). Figure~\ref{fig:ucell} (c)  
shows how these two layers are spatially arranged one with another.}
This now seems to be clear that the commensurability of the (110)
face of cubic diamond and the hexagonal lattice has the most significant
impact on the selective formation of the cubic diamond network. \\

\noindent \textbf{Removing external field} \\

\begin{figure*}
  \centering
    \includegraphics[width=\textwidth]{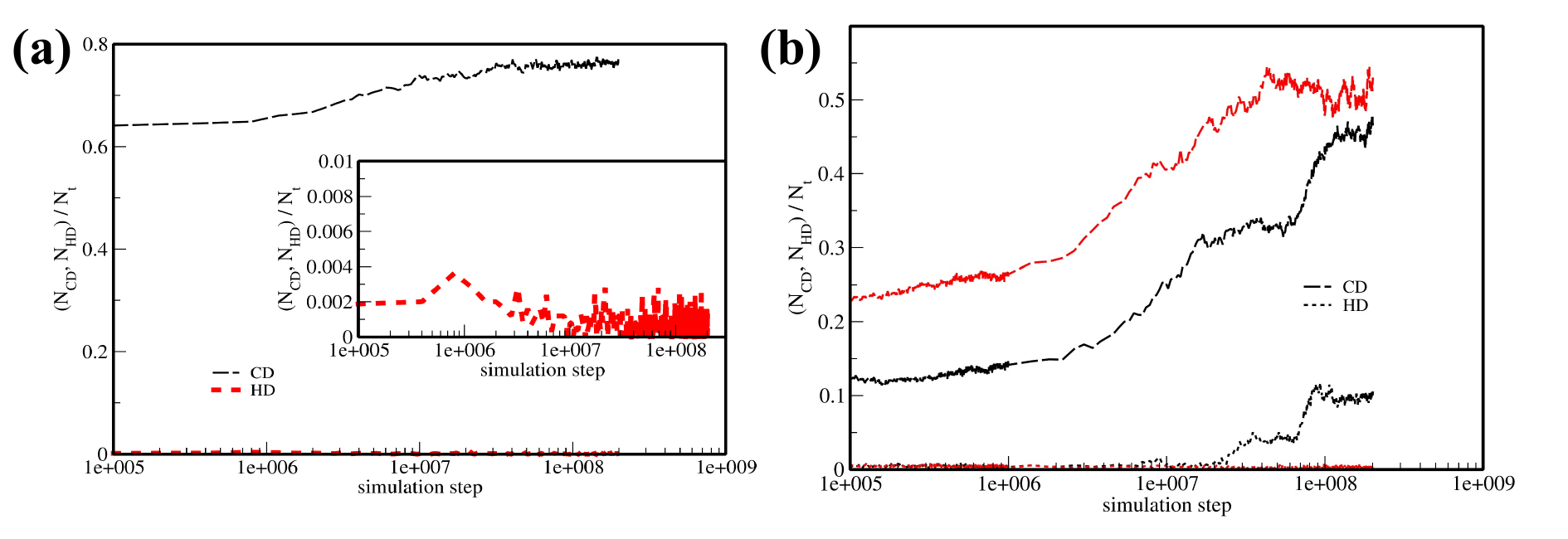}
    \caption{
     Parts (a,b): The time evolution of the number of patchy particles
     belonging to CD and HD networks when the external field is removed
     for the following conditions,
     (a) $\xi=1.0$, $T^*=0.29$; 
     (b) $\xi=1.0$ and starting configuration taken from 
     $T^*=0.33$ (black lines) or $T^*=0.325$ (red lines)
     and cooled to $T^*=0.31$.}
     \label{fig:removing}
\end{figure*}

In the previous section, we demonstrated that increasing  
the external field results in the selective emergence of the cubic diamond 
network that is compatible with the structure of the first adlayer.
Therefore, a naturally arising question is, what happens when the external field is turned off? Figure~\ref{fig:removing} demonstrates the absolute ratio of the number of molecules
belonging to either of the diamond phases ($N_{CD}$ or $N_{HD}$) to the total
number of patchy particles in the system for several values of the 
parameter $\xi$. For $\xi=1$ (Fig.~\ref{fig:removing} (a)),
we see that once over 60\% of the system crystallized into cubic diamond,
the removal of the external field causes further growth of the formed phase. 
The same concerns $\xi=0.6$ and $\xi >1$ cases (Supporting Information). 
For $\xi=0.4$ the situation is slightly different as the
removal of an external field does not change anything since
the system behaves as a bulk-like (type-3 growth mechanism) irrespectively (Supporting Information). 

However, the above refers to the low-temperature configurations with 
quite a large nuclei in the range of the bulk phase diagram,
where such crystal networks are stable. 
Therefore, for $\xi=1$ we checked what would happen 
once the nucleus is comprised of several hundred atoms. 
Since the phase diagram terminates above $T^*=0.315$ 
with a (metastable) critical point, we took the configurations
from higher temperatures, i.e. $T^*=0.33$ and $T^*=0.325$ 
and cooled it down to $T^*=0.31$, so within the bulk stability region, and results are shown in Figure~\ref{fig:removing} (b). 
In the former, the initial nucleus of a cubic diamond 
comprised of merely 150 tetrahedral patchy particles 
grows further, however after enough time, the formation
of a competitive hexagonal diamond can be seen.
In the second case, the initial nucleus was larger and composed of 
around 400 patchy particles and the defects are not present 
within the considered simulation time. 

These findings are crucial for the following reasons. 
The most important is that the 
application of the confining potential can be used to tune 
the architecture of the self-assembled networks in the system comprised 
of tetrahedral patchy particles which remain stable even if 
the external field is turned off.
Moreover, we demonstrated that the external field can be used
as an initial tool to increase the density next to the surface
and to create an initial nucleus that, if large enough, continues to grow 
into the same network. 
Both of these aspects open up an avenue 
for the post-synthetic treatment of these colloidal diamond crystals
that can be potentially useful for photonic applications. \\

\noindent \textbf{System size effects} \\

\begin{figure*}[t!]
    \centering
    \includegraphics[width=\textwidth]{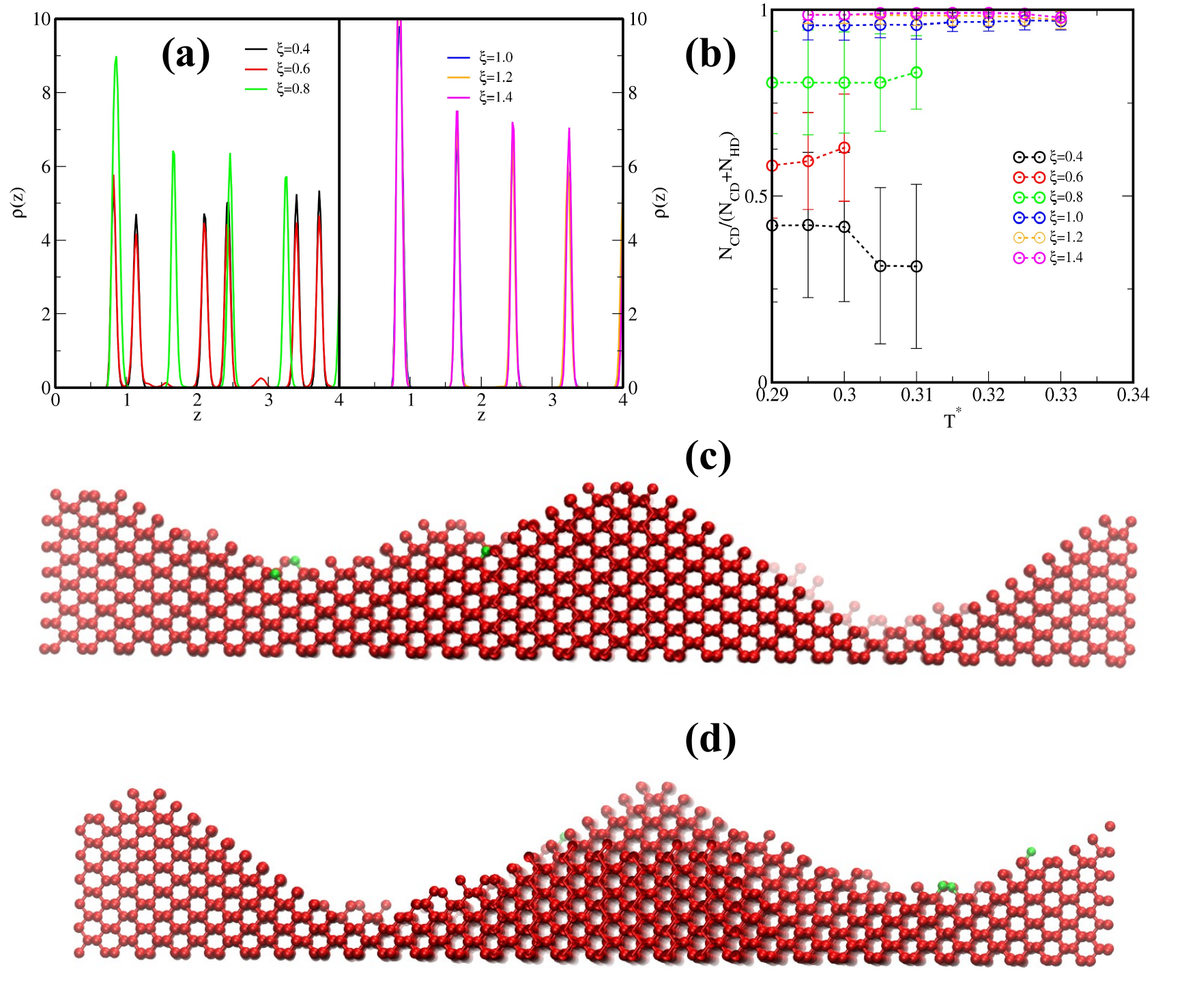}
    \caption{Part (a): Density profiles for considered systems under different
strengths of an applied external field. Part (b): temperature relation of
the cubicity parameter for all considered systems.
Parts (c), (d): Snapshots for $\xi=1.0$ and $\xi=1.2$
at $T^*=0.3$}
    \label{fig:big}
\end{figure*}

As the nucleation process is an activated event, 
we see that despite being able to selectively
obtain the cubic diamond for most of the replicas, 
in some cases, the growing crystal is tilted
by $45^\circ$ relative to the surface,
exposing the (111) face of a cubic diamond and 
giving rise to the formation of stacking hybrids.  
We envisage, that since our studies are performed in $NVT$ ensemble,
the lateral directions are not able to fluctuate and adjust according
to the lattice parameters of the forming crystals. In consequence,
frustrations emerge, and the resulting imperfections 
may affect the growth process.

To verify our hypothesis, we performed auxiliary simulations 
for system sizes comprising of $16129$ designer patchy particles,
corresponding to approximately seven times larger surface area.

The results are qualitatively identical to those presented for
a smaller system consisting of 2500 particles and can be
found in Fig.~\ref{fig:big}. 
The density profiles displayed in part (a) 
are identical to those for a smaller system. 
However, as expected, the frustrations that were present in
the smaller system vanish which is demonstrated by the values in
the cubicity parameter (cf. Fig.~\ref{fig:big} (b) and Fig.~\ref{fig:external} (b))
for systems with $\xi \ge 1.0$. 
It is evident that these systems are comprised of a cubic diamond crystal
over an entire temperature range. Within all five replicas, we did 
not spot any significant defects arising from the 
emergence of competing hexagonal diamond. 

\begin{figure*}[t!]
 \centering
    \includegraphics[width=\textwidth]{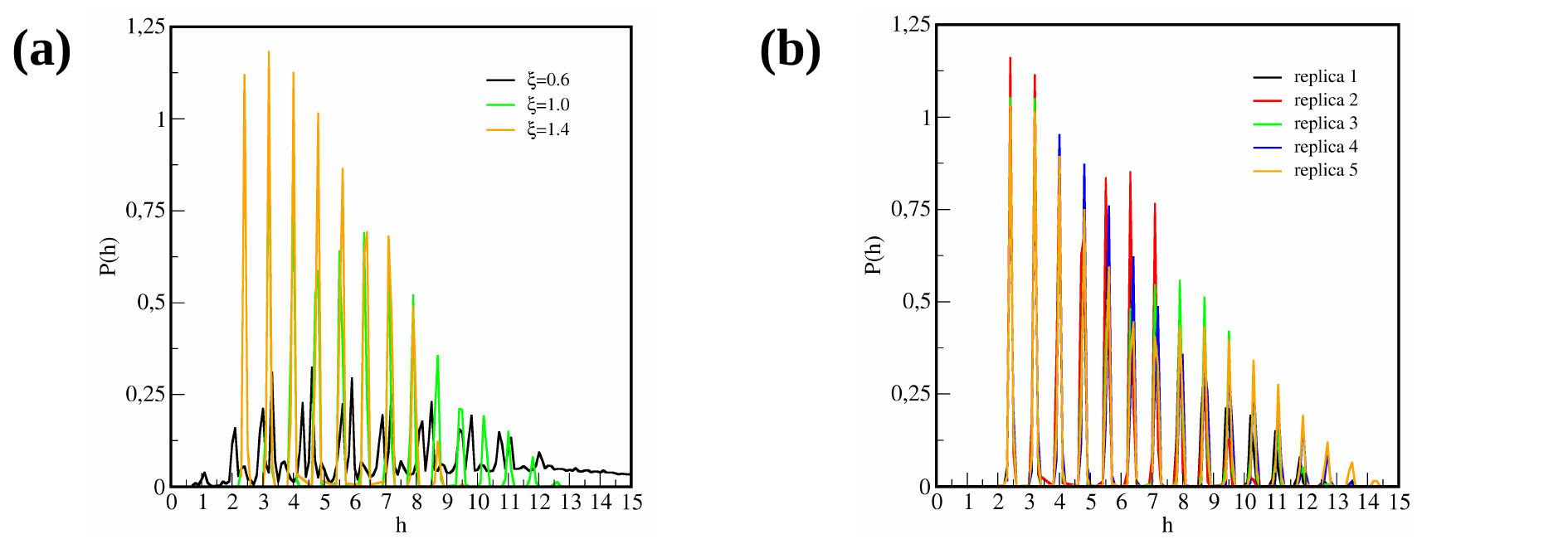}
    \caption{Part (a): Distribution of the film height for different strength of the surface potential.
    Part (b): Distribution of the film height for different replicas for the system $\xi=1.0$ and $T^*=0.30$.}
    \label{fig:height}
\end{figure*}

Another aspect worth highlighting regards the type-1 growth
mechanism (cf. Fig~\ref{fig:external} (c))
whose emergence was not revealed due to surface fluctuations
which  are neglected in the calculation of the average excess adsorption.
Snapshots of large systems for $\xi=1.0$ and $\xi=1.2$
seem to corroborate that assumption where one can clearly see a large 
differences in the film height in the xy-plane. 
To demonstrate that, we calculated the distribution of the film height $P(h)$
which is shown in Figure~\ref{fig:height}. 
Part (a) display the comparison of the distribution $P(h)$
for different strengths of the surface potential. 
We can see that these systems, although having the value of 
excess adsorption of $\Gamma \approx 6$ (cf. Fig.~\ref{fig:external} (c)),
are significantly different from one another. 
System for $\xi=0.6$ exhibit the largest film height, as expected
due to the type-1 growth mechanism. With the increase
of the strength of confining potential, film height 
decreases, indicating that the layers tend to 
grow more uniformly, in a layer-by-layer fashion (type-2 or type-3 mechanisms). 
Part (b) on the other hand, shows results for the system $\xi=1.0$ and 
demonstrate that even for such large system sizes, we still observe 
the differences between distinct replicas.

Moreover, this is particularly interesting since the formed 
cubic crystals expose their (111) face and unlike in the small system,
the hexagonal polymorph does not grow. This observation 
supports our previous suspicion that the emergence of a competitive
crystalline network originates from frustration due to insufficient
surface size which on top of that cannot fluctuate in $NVT$ simulations.
Now it seems to be clear that large enough system sizes 
alleviate the issues encountered by insufficiently large systems. 
This finding is crucial and demonstrates the importance of finite-size effects in computer simulations and the need for using multiple replicas in order
to attain representative results.

\section{Conclusions}
In this paper, we performed a molecular dynamics simulation study
demonstrating the possibility of the selective formation of the cubic diamond lattice.
The process relies on the use of an external field which, if strong enough, 
causes the adsorption of the first adlayer that is commensurate with the (110) face of a cubic diamond. 
Such a result is of paramount importance as the cubic polymorph 
of the diamond-family networks exhibits the complete photonic bandgap
making it potentially useful in terms of photonic crystal applications. 

In order to comprehensively assess the influence of the applied
external field, we evaluated the bulk phase diagram which is of 
similar type to the one reported for tetrahedral patchy particles
modelled with the Kern-Frenkel potential. 
This suggests that despite using a different approach, our study
should not be model-dependent and demonstrate a universal behavior. 
The latter is strongly supported
due to the fact that our findings are based mainly on the geometric commensurability of the two networks.
On top of that it was only recently shown
that metastable cubic ice can be obtained due to heterogeneous nucleation for an atomistic
water model \cite{IceIc_Michaelides}.

Strikingly, once the external field is removed, the formed
networks remain stable. This indicates the possibility of further post-synthetic use in which the cubic diamond network could be "washed out" from the surface. Such a procedure
is a common practice for the synthesis of 2D materials. 
Such a finding opens up an avenue for experimental approaches aimed at 
utilization of tetrahedral patchy particles that not only 
selectively assemble into a cubic diamond crystal but also can be 
tested for their photonic applications and specific post-synthetic
adjustments can be made. 

Eventually, we highlight the importance of using multiple replicas to 
demonstrate statistically relevant results. On top of that,
it is evident that in such systems, the results are subject to strong
finite-size effects so the necessity to use sufficiently large systems 
is crucial.  

\section*{Conflicts of interest}
There are no conflicts to declare.

\section*{Acknowledgments}
We would like to thank Andrzej Patrykiejew and Ma{\l}gorzata Bor{\'o}wko
for careful reading of the manuscript and for providing us helpful comments on this article. 
This work was supported by the National Science Centre, Poland, under Grant No. 2021/41/N/ST4/00437, PRELUDIUM 20.

\bibliography{biblio}
\bibliographystyle{rsc}

\clearpage



\section*{Supplementary material}

# Pursuing Colloidal Diamond
# Supporting Information

Łukasz Baran, Dariusz Tarasewicz, Daniel M. Kamiński, Wojciech Rżysko

# 1 Justification of the chosen model

Model used throughout the current study, was similar as already reported in our previous papers [1, 2]. We found that the choice of the model parameters to be $\sigma_a = 0.2\sigma$, $\varepsilon_{aa} = 5.0\varepsilon$, and $r_{cut} = 2\sigma_a$ is the most efficient to describe relatively strong but reversible associative interactions. The parameter $l$ allows one to the precise control over the desired patch's valency. In order to study diamond phases, we found that $l = 0.36\sigma$ allows for single association of each active sites, resulting in the formation of the total of four bonds per particle. To demonstrate that, Figure S1 displays the distribution of a number of nearest neighbors $N_{nn}$ per particle for three different embedment distances being $l = 0.36\sigma$, $l = 0.40\sigma$, and $l = 0.45\sigma$, calculated up to the first minimum in the radial distribution function $r_{max} = 1.2\sigma$.

One can see that $N_{nn} = 4$ dominates for $l = 0.36\sigma$ and $l = 0.40\sigma$, however, in the case of the latter we see that it can take values between $5 - 7$ indicating the possibility that a patch can have higher valency than one. Systems with $l = 0.45\sigma$ take values of $N_{nn} = 9, 12$ as the most probable, indicating the possibility for the formation of fcc/hcp-like crystalline networks which was beyond the scope of current study.

# 2 Bulk phase diagram

As mentioned in the main text, the phase diagram has been evaluated using block's analysis method proposed by Binder [3]. The distribution functions for several distinct densities and different temperatures are shown in Fig. S2. Part (a) display a regular distribution exhibiting double-peak behavior corresponding to the two-phase coexistence. In present case, we are dealing with the gas-solid coexistence, the latter phase being an hybrid of interwoven hexagonal and cubic diamond polymorphs at around $\rho = 0.71$.

In part (b), on the other hand, one can see that at temperature $T^* = 0.305$ and the density $\rho = 0.4$, there is a gaussian peak, indicating disordered fluid, that is uniformly distributed over an entire system space. This means that in such conditions, we are above the transition temperature into the ordered phase. The same concerns higher densities, however, it needs to be emphasized that suitable transition temperature accordingly increases.

In part (c), we observe large peak around $\rho = 0.71$ corresponding to diamond phases as previously. On the other hand, the low-density peaks vanished which corroborate that gas-solid coexistence is not present under such circumstances. Strikingly, other peaks with slightly lower density than the crystalline phase appear. We conjecture that this is due to the presence of a liquid phase that emerges in a relatively narrow region. Moreover, we did not observe fluctuations between gas and liquid phase within this range of temperatures and for all densities studied, therefore, we envisage this liquid phase is metastable.

In part (d), at higher density, we observe the coexistence between diamond phases and BCC, the latter of which is composed of two interwoven diamond phases. Presence of BCC and other phases has been evaluated by means of the calculation Steinhard order parameter (Eq. 2 and 3 in the main text). The criteria for the discrimation of particular crystalline environments can be found in Table I of the following article [4]. Corresponding snapshots for densities

$\rho = 1.0$ and $\rho = 1.2$ are displayed in Figure S3. In view of all the above, this phase diagram is of a similar type as the ones reported by Romano *et al.* [5, 6], although performed for different interparticle potential.

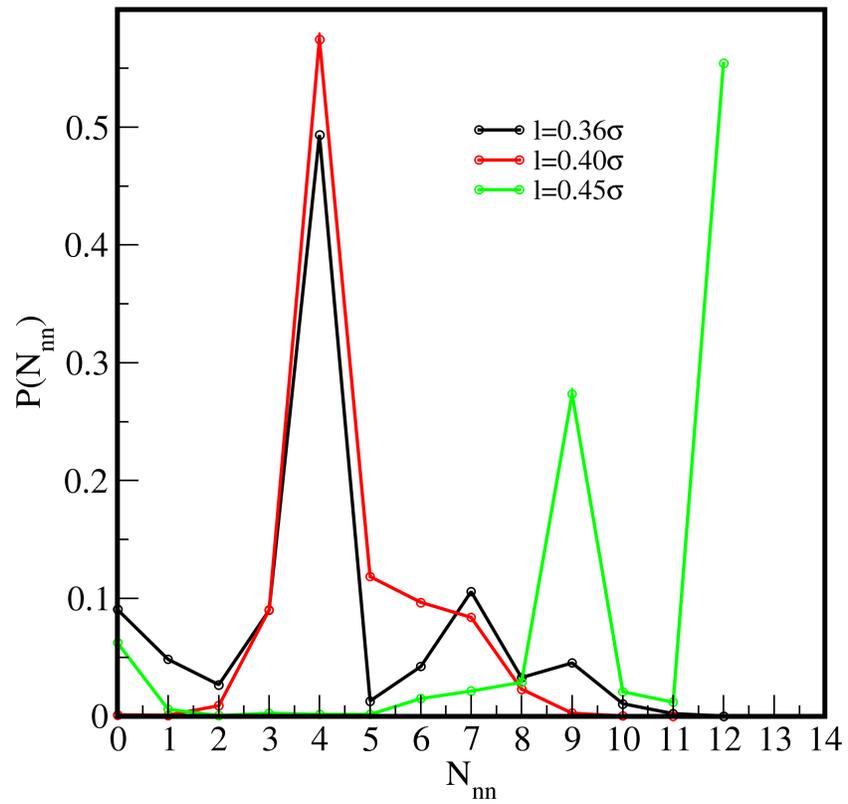

Figure S1: The distribution of the nearest neighbors for systems differing in the embedment distance $l$.

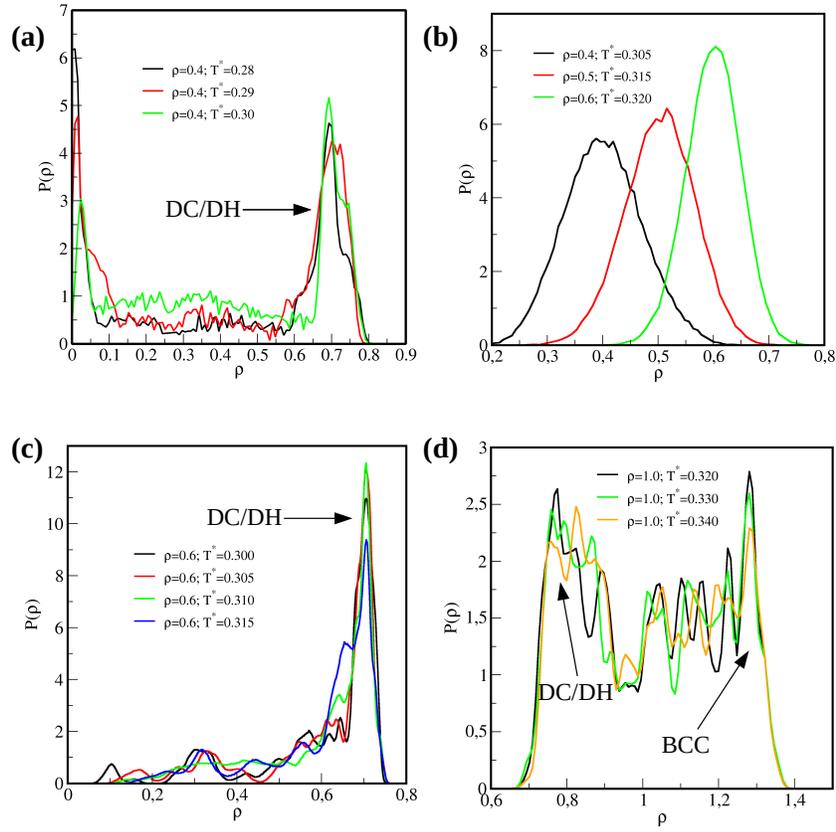

Figure S2: The density distribution functions of the largest block for bulk systems at different densities and temperatures.

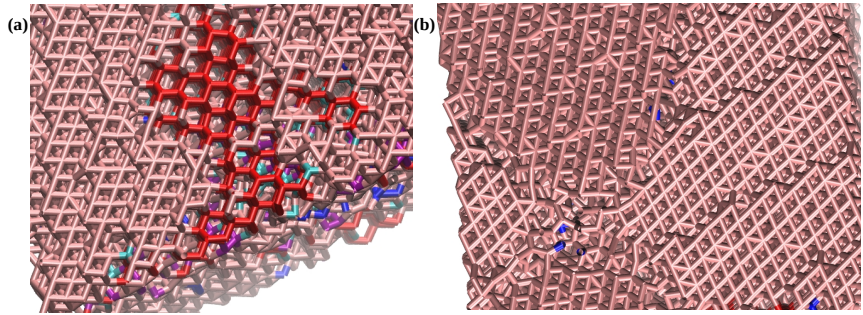

Figure S3: Fragment of the configurations for systems in $\rho = 1.0$ (a) and $\rho = 1.2$ (b) at $T = 0.34$. Red sticks correspond to cubic diamond whereas pink sticks is the remaining fluid.

5

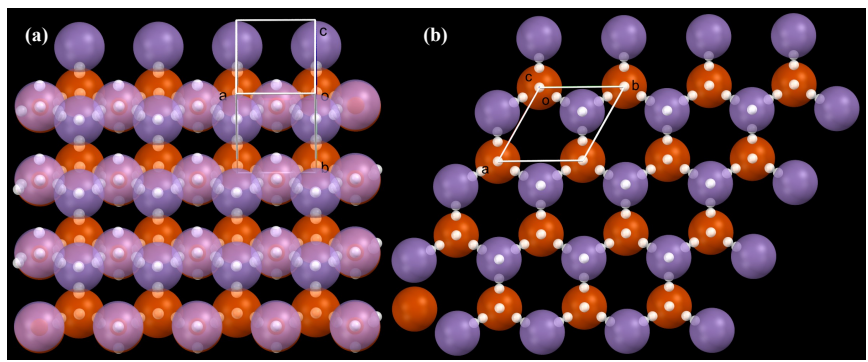

Figure S4: Fragment of the configurations of (a): (110) and (b):(111) faces of cubic diamond formed by tetrahedral patchy particles for systems $\xi = 1.0$ (a) and $\xi = 0.4$ (b). Yellow and purple atoms correspond to the lower and upper layer, respectively. White spheres schematically represent the active sites.

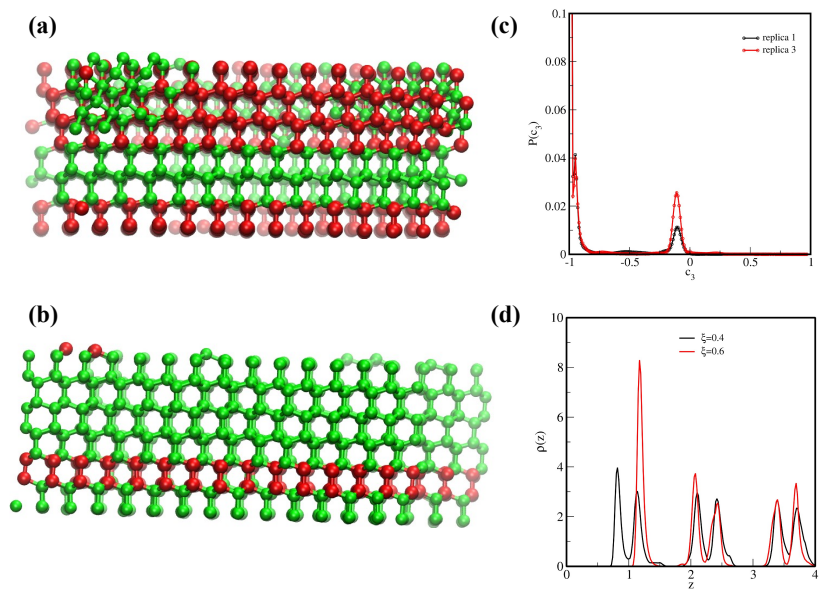

Figure S5: Parts (a, b): Snapshots of two different replicas, differing in the ratio of two cubic polymorphs in the stacking hybrids. Part (c): Distribution function of the order parameter $P(c_3)$ for replicas shown in parts (a, b). Part (d): Comparison of density profiles for $\xi = 0.4$ and $\xi = 0.6$. The latter is shifted by $1/3\sigma$ in the z-direction (cf. Main text).

6

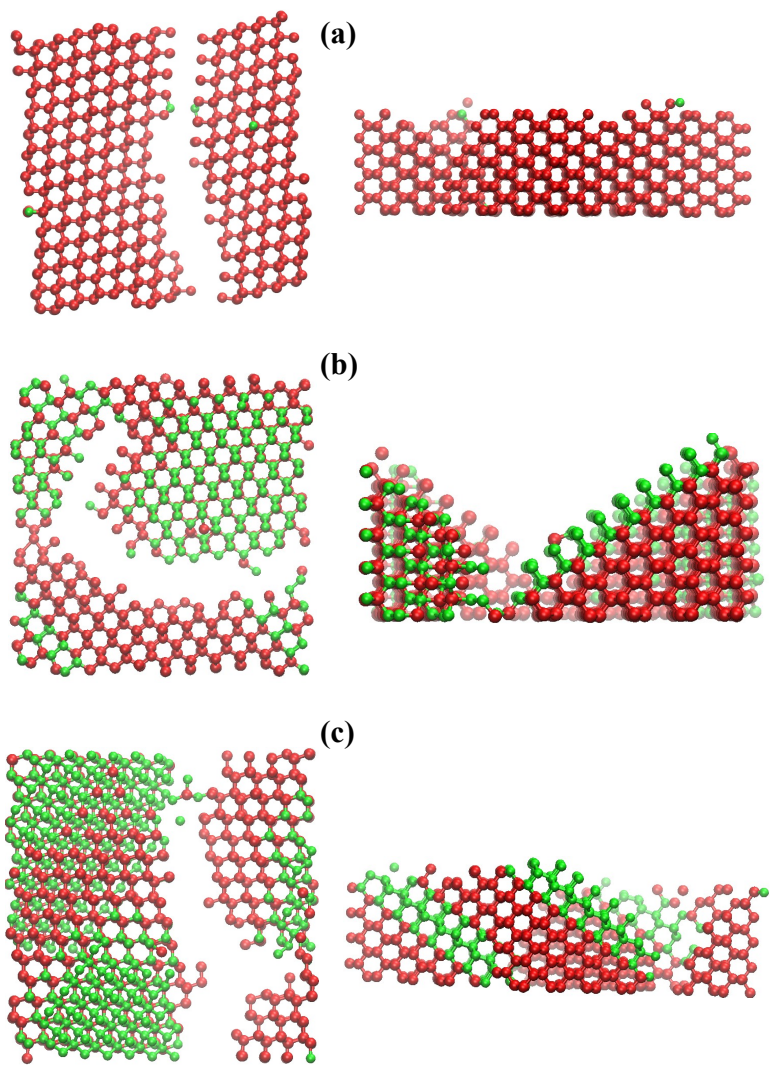

Figure S6: Snapshots for three different replicas for system $\xi = 1.0$ at $T^* = 0.30$. Left and right panels correspond to top and front view, respectively.

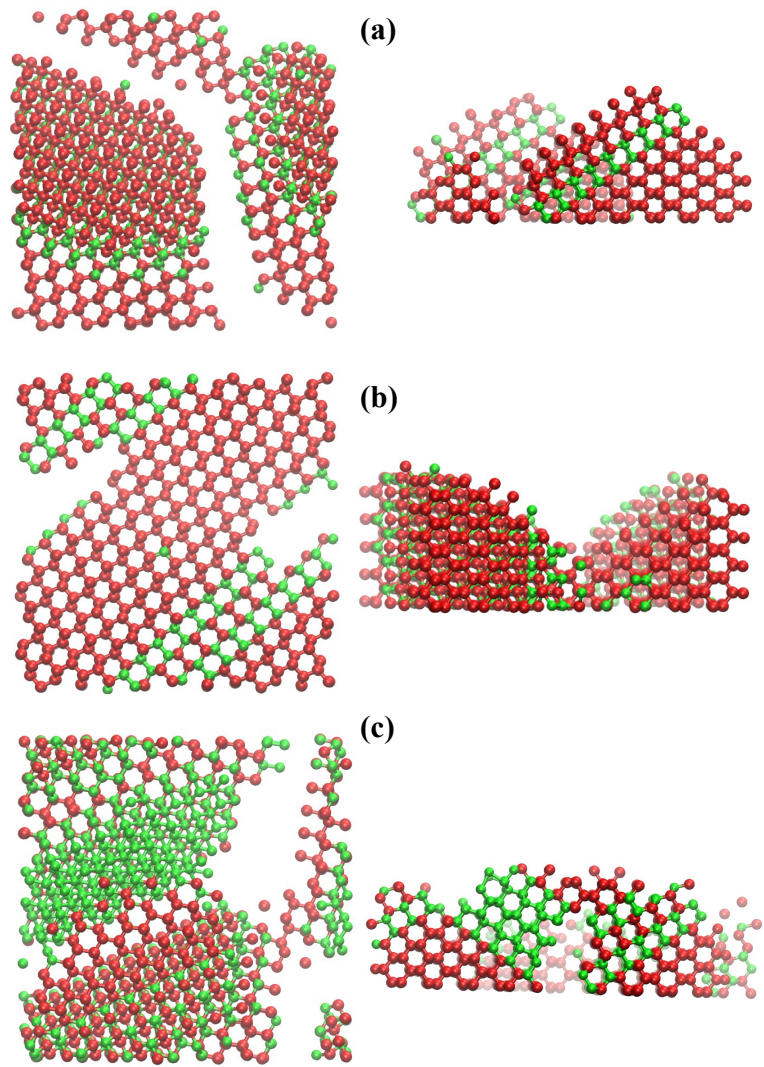

Figure S7: Snapshots for three different replicas for system $\xi = 1.2$ at $T^* = 0.30$. Left and right panels correspond to top and front view, respectively.

8

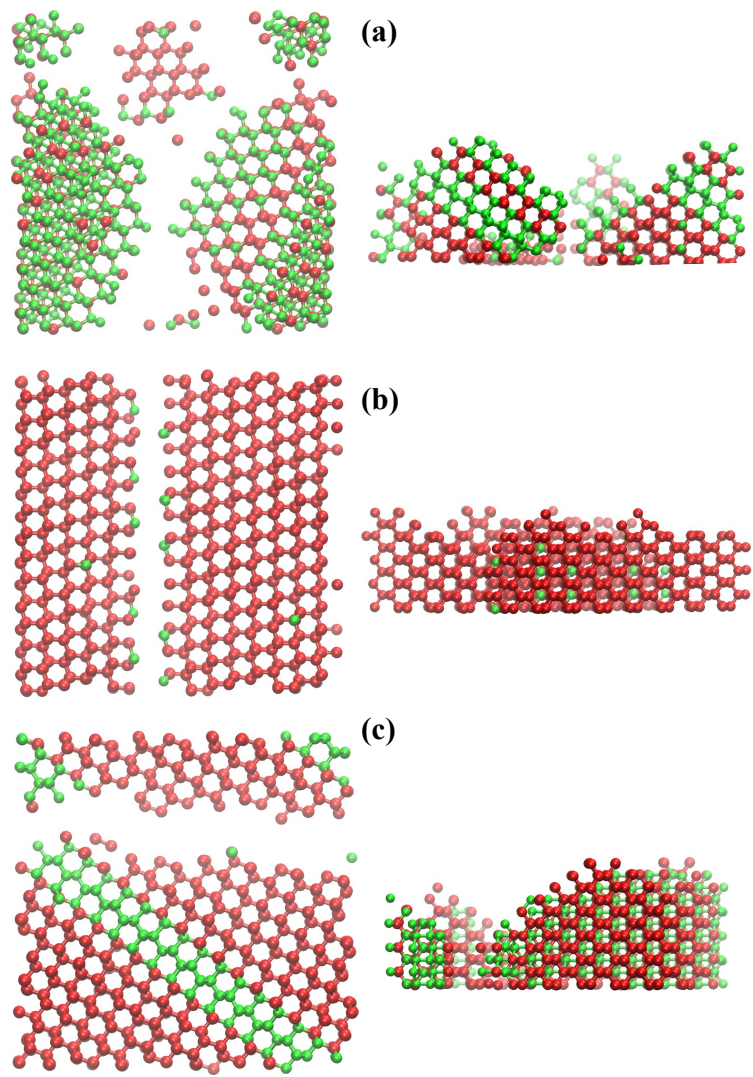

Figure S8: Snapshots for three different replicas for system $\xi = 1.4$ at $T^* = 0.30$. Left and right panels correspond to top and front view, respectively.

9

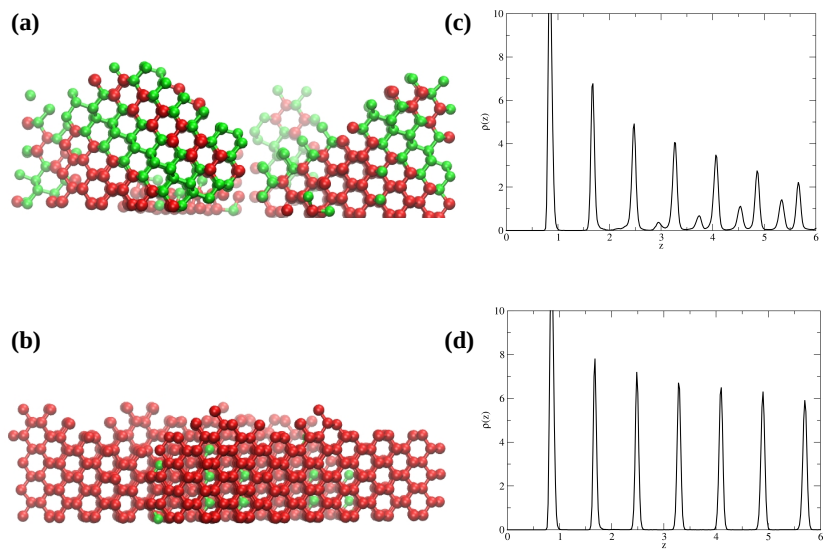

Figure S9: Parts (a, b): Snapshots for two different replicas for system $\xi = 1.4$ at $T^* = 0.30$. Parts (c, d): Density profiles for replicas shown in parts (a, b) with tilted (c) and flat growth of the crystal (d).

10

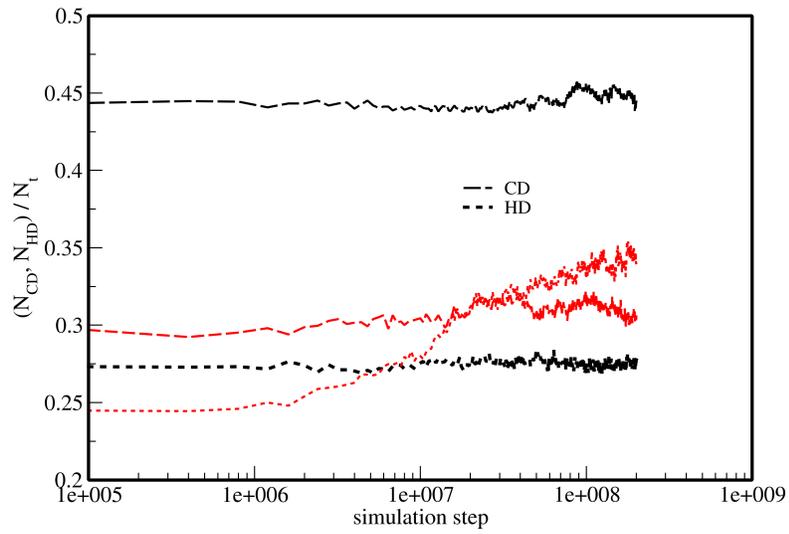

Figure S10: The time evolution of the number of patchy particles belonging to CD and HD networks when the external field is removed for $\xi = 0.4$, $T^* = 0.29$ (black lines) and $\xi = 0.6$, $T^* = 0.29$ (red lines)



\end{document}